# Anonymization of Whole Slide Images in Histopathology for Research and Education

*Authors*:

Tom Bisson*[1], Michael Franz*[1], Isil Dogan O[1], Daniel Romberg[2], Christoph Jansen[1], Peter Hufnagl[1,3], Norman Zerbe[1]

1 Charité – Universitätsmedizin Berlin, corporate member of Freie Universität Berlin and Humboldt-Universität zu Berlin, Institute of Pathology, Charitéplatz 1, 10117 Berlin, Germany

2 Fraunhofer Institute for Digital Medicine MEVIS, Am Fallturm 1, 28359, Bremen, Germany

3 University of Applied Sciences (HTW) Berlin, Centrum für Biomedizinische Bild- und Informationsverarbeitung (CBMI), Ostendstraße 25, 12459 Berlin, Germany

* Contributed equally

# Abstract

**Objective:** The exchange of health-related data is subject to regional laws and regulations, such as the General Data Protection Regulation (GDPR) in the EU or the Health Insurance Portability and Accountability Act (HIPAA) in the United States, resulting in non-trivial challenges for researchers and educators when working with these data. In pathology, the digitization of diagnostic tissue samples inevitably generates identifying data that can consist of sensitive but also acquisition-related information stored in vendor-specific file formats. Distribution and off-clinical use of these Whole Slide Images (WSI) is usually done in these formats, as an industry-wide standardization such as DICOM is yet only tentatively adopted and slide scanner vendors currently do not provide anonymization functionality.

**Methods**: We developed a guideline for the proper handling of histopathological image data particularly for research and education with regard to the GDPR. In this context, we evaluated existing anonymization methods and examined proprietary format specifications to identify all sensitive information for the most common WSI formats. This work results in a software library that enables GDPR-compliant anonymization of WSIs while preserving the native formats.

**Results**: Based on the analysis of proprietary formats, all occurrences of sensitive information were identified for file formats frequently used in clinical routine, and finally, an open-source programming library with an executable CLI-tool and wrappers for different programming languages was developed.

**Conclusions**: Our analysis showed that there is no straightforward software solution to anonymize WSIs in a GDPR-compliant way while maintaining the data format. We closed this gap with our extensible open-source library that works instantaneously and offline.

# Introduction

**Motivation**

Digital processing of medical image data has experienced a rapid increase in importance for modern diagnostics over the last decades. While radiology went digital decades ago, the digitization of pathology is still in its infancy. The reason for this technical deficit is inherent to the field, as radiology has been a purely digital domain since its beginning. In pathology, on the other hand, a large part of the work (e.g., biopsy, preparation, sectioning) is still performed in analog form, so digital analysis methods are only used to a limited extent. Additional reasons for the slow adoption of digitization are the significantly larger data volumes and computing capacities required to process histopathological images. In the early days of research in digital pathology, topics such as image data acquisition and visualization played a major role. Other topics are now taking center stage, and they include standardization, web-based diagnostics and data mining for the development of artificial intelligence (AI) solutions, such as machine learning (ML) or neuronal networks [1, 2]. Since all these issues involve the exchange of image data, the risk of disclosing sensitive patient-related data increases. In digital pathology, high-resolution scans of tissue on glass object slides, known as Whole Slide Images (WSI), are most commonly used. The scanner device market is fragmented among multiple manufacturers introducing a variety of proprietary WSI file formats, which may differ significantly depending on the scanner type and software version. Depending on the file format, sensitive information may be stored in different locations and extents. In clinical routine, each case has its own case ID, which is usually printed on the slide either as plain text or as a 1D or 2D barcode used by physicians to identify the individuals being examined. When scanning, the label with the ID is always captured and stored as an image file in the WSI. Additionally, some scanners can digitize slide labels and store them within the file format, among other acquisition-related metadata. Consequently, when exchanging WSIs, strict attention should be paid to whether the partner receiving the data is permitted to access patient-related metadata or prohibited from it. For example, while billing for second opinions requires this data, there is generally little need for the sensitive information stored in WSIs for educational purposes or research contributions. Regardless of the partners involved in the data exchange, the transmission channel and networking security also have a major impact on data protection. For these reasons, it is crucial to be able to anonymize WSIs. There are two approaches to ensure that no sensitive data leaks during the exchange. First, the interaction with the image data is separated from the identifying information. An access control system could be implemented, for example, via a web service or a programming interface to achieve this separation. This measure would ensure that end users or applications never have direct access to the WSI and, thus, to sensitive information. The data could still be made available using appropriate authentication, so that use cases such as billing for second opinions would still be possible. The second approach is to remove all sensitive information directly from the respective WSI. The case ID could no longer be read from the respective file, but in return, the WSI can be shared without having to implement and host appropriate access control. This article presents an anonymization policy and a concrete solution for the second approach, which supports a wide range of data formats from different slide scanner manufacturers.

**Regulatory Aspects**

Health-related data are generally subject to special protection and conscientious treatment of such data is crucial from an ethical and legal perspective. These data may contain information that allows conclusions about the patient's circumstances or state of health. In Europe, the fundamental handling of person-identifying data is governed by the European General Data Protection Regulation (GDPR), which went into effect in 2018 and defined a legal framework as a minimum requirement for all EU countries. No specific guidelines are provided; instead, it states that the protection of personal data must ensure a 'reasonable' level, although the wording is very open to interpretation [3]. Apart from the fact that health-related and medical data are not explicitly addressed in the GDPR, it does contain rules for the processing, storing and sharing of any data identifying a natural person [4]. Improper handling of such data can be imposed with massive fines [5]. Regarding the health sector, most penalties are imposed for inappropriate or insecure handling of patient data. These penalties are comparatively low for hospitals and medical institutions, as they are not intended to drive them into bankruptcy but rather to motivate them to handle health data more responsibly and to close security gaps within their data retention and provision.

In the United States, the handling of health information is covered by the stricter Health Insurance Portability and Accountability Act (HIPAA) of 1996, which governs any health-related information that contains data identifying a patient. Four categories of violations exist, carrying potential penalties with a maximum fine of 1.5 million US dollars per year in the highest category [6]. In contrast to the US, other countries (e.g., Brazil and Japan) have laws similar to the GDPR or orientate towards these regularities [7].

**Anonymization of medical image data**

One way to prevent careless handling of medical data is to delete all patient-identifying records. Such anonymization can be interpreted quite openly concerning the vague formulations of the General Data Protection Regulation. To evaluate anonymization in the context of the GDPR, Vokinger et al. proposed a reference classification scheme for different levels of anonymization of medical data [8]. They distinguish between fully identifiable data, reversibly and irreversibly anonymized data and data that is anonymous *per se*. Only when a data record is irreversibly anonymized does it cease to fall within the scope of the GDPR, allowing it to be further disseminated without legal or regulatory concerns. However, it is practically impossible to anonymize WSIs irreversibly, as the digitized tissue itself can serve as an identifier if both the original tissue and its association with the patient are accessible. However, the GDPR already considers data anonymous if it can only be traced back to specific individuals with a disproportionate effort. In the further course of this paper, anonymization, therefore, always refers to irreversible anonymization. Most WSIs are fully identifiable via a case ID, as pathologists and doctors require the patient's data to make a diagnosis and derive recommendations. In the research and educational context, all sensitive information must be removed to ensure unauthorized third parties cannot infer the patient's condition. In publications, WSI anonymization is often performed by the physical covering of the slide label before scanning, which prevents the patient identification number from being captured by the scanning device [9, 10, 11]. While this method is very simple, it is, however, only viable if access to patient-related information is not intended at all. This method is not a practical solution for off-label use of WSIs from clinical routine (e.g., for research and education), as this would require a second scan iteration. Besides storing the label image, additional metadata can be found in the file formats. For example, modern scanning devices can read out barcodes printed on the slide label and store them in a uniform encoded way. Another source of identifying data is information about the acquisition of the WSIs, such as scanner IDs, users and dates that can potentially induce the traceability of clinical cases. For more details on metadata, see the Materials and Methods section.

**Related Work**

Two different approaches for anonymizing WSIs can be identified in the literature and from practical use cases. One approach is to remove all sensitive information from the respective WSI while retaining the original data format. The second approach involves first converting the WSIs into a file format that can either be anonymized with little effort or is anonymous *per se*.

The latter is always the case when the data format contains only the image data of the tissue and therefore lacks metadata or associated images. For example, there is the possibility of converting WSIs into the DICOM WSI format. DICOM is a standard for storing medical image data, in which additional metadata can be stored for both the patient and the imaging conditions, as well as any computations and analyses that may have already been performed. DICOM has been established in radiology for decades and thus, there are well-tested approaches for anonymizing DICOM datasets. In fact, Working Group 18 defined a comprehensive list of DICOM attributes used in clinical studies in Supplements 70 and 142, which serve as a conceptual framework for the underlying anonymization process [12, 13]. While it is expected that the DICOM WSI standard will become widely adopted in pathology, until then, a solution for anonymizing WSI in existing proprietary formats is needed. Also, with the Open Microscopy Environment (OMERO), anonymization of WSIs can be implemented. If these are converted to the generic ome.tif format, only the image data of the tissue is transferred, but not the metadata and any additionally stored images. The converted WSIs can then be read with the BioFormats library [14]. As OMERO is a community-driven open-source project, the further development and adoption of the format and the associated library are strongly dependent on the acceptance of the developer and user community. For anonymization in both DICOM and OMERO, the respective WSIs would first have to be converted to another format. Besides requiring time and effort, this is not feasible if the tools in use cannot handle the target format.

An open-source Python script to anonymize WSIs for file formats from 3DHistech, Aperio and Hamamatsu was published on GitHub in 2014 [15]. This script provides rather limited anonymization as it solely removes the macro and label image from the WSI. In addition, the tool is severely limited to outdated software versions of the scanning device, as it is no longer actively developed and contributions from the GitHub community are not accepted by the author. An alternative open-source software is the *svs-deidentifier* [16] which is forked from the python script described beforehand but can be operated with a graphical user interface via the browser. The tool supports the newer .svs format originating from the Leica GT450 but lacks support for the common WSI formats from other vendors. 3DHISTECH provides the slide converter, which anonymizes WSIs in the vendor's proprietary Mirax file formats. Schüffler et al. report their custom solution for iSyntax files based on these scripts but do not make it available as open-source software [17].

At present, there is no open-source solution that supports the heterogeneous range of current WSI formats. Although there are existing tools that can initially anonymize a small selection of formats, they do not provide the possibility to extend them to new or updated image formats and only allow a rather weak anonymization.

**Contributions**

With the ever-growing importance of cloud-based services in digital pathology, the need to anonymize WSIs is also becoming increasingly important. These web services should provide the means to guarantee that no sensitive data is transmitted during the upload of WSIs. Only then can the protection of patient data be ensured and thus be realized following the requirements of the GDPR. To facilitate the implementation and evaluation of such anonymization, we propose an anonymization policy in which we categorize already existing and yet-to-be-developed options for anonymizing WSIs regarding the data protection and the regulatory requirements of the GDPR. Although DICOM WSI already implements all the necessary functionality for strong anonymization, it does not play a role in practical everyday life as it is not fully supported in image management systems or scanning software yet. It can therefore be assumed that the anonymization of WSIs in the context of research and teaching will have to operate without DICOM for several more years. Furthermore, we present a vendor-agnostic software library to anonymize various proprietary file formats accordingly. In addition, we propose a solution for web-hosted services, where anonymization is integrated directly into an upload process to prevent potential GDPR violations.

## Methods

**Anonymization Policy**

The comprehensive removal of sensitive information first requires detecting potentially identifying data within each of the proprietary WSI file formats. This is particularly important because there may be data that is not further critical on its own but, in combination with other non-critical data, may, in turn, allow the patient to be identified. For example, if the case ID is deleted in a WSI, the combination of image data, acquisition date and a scanner's serial number can make re-identification possible. We have identified the following four key elements as sources of confidential information inside WSIs:

- Filenames
- Associated images
- Acquisition and image metadata
- Tissue image data

Filenames may contain identifying information such as the case ID, study name, or indication. Addressing this, however, requires no significant effort and will therefore not be considered any further in the following. Typically, WSIs contain associated images that show the slide label and a macro image of the physical glass slide (as shown in Figure 1) that were acquired during the digitization process. Depending on the file format, the macro image may contain the entire slide, either with or without the label, or only the relevant part of the tissue. In some formats, bounding boxes are additionally drawn to indicate the scanned regions. The label is not always stored in a separate image. WSIs from Hamamatsu's Nanozoomer, for instance, only contain a macro image of the entire glass slide.

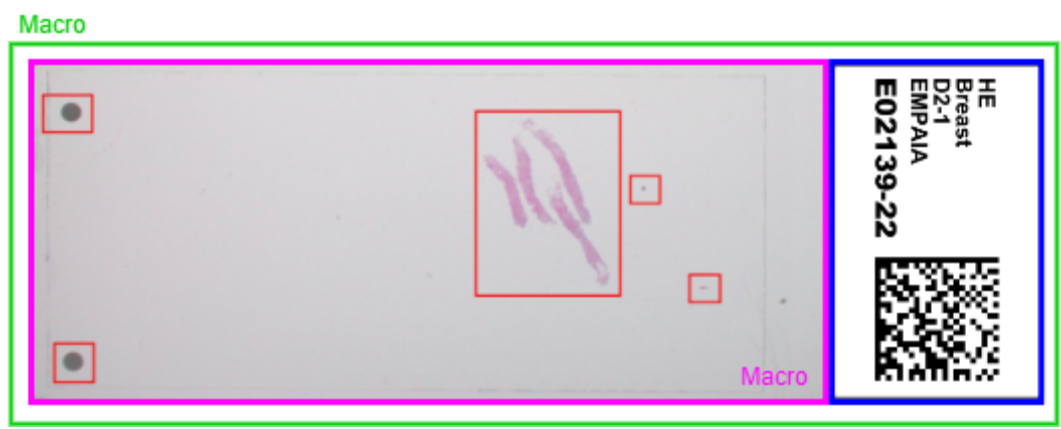

**Figure 1**: Histologic glass slide: The blue frame shows the slide label, while the green and purple frames show different representations of the macro image, depending on the scanner implementation. The red frames indicate the bounding boxes of the scanned regions within the macro image.

Detailed information about the structural design and attribute positions were gathered through file format specifications, examinations of scanning and viewing software and dialogues with scanner vendors. We then grouped all metadata as to whether it is related to a human subject, such as the case ID, or to acquisition modalities, such as scanner IDs, serial numbers, or date and time of recording. Appendix A provides a complete overview of the categorized metadata for each file format. We have defined an anonymization policy that discriminates between five different anonymization levels for WSIs. Figure 2 shows the extent of data protection and GDPR compliance at each level and the current availability of corresponding solutions. The higher the anonymization level the stronger the data protection security achieved. Levels I and II correspond to the first attempts to remove sensitive data from WSIs by removing real names and identifiers from filenames and dereferencing associated images. By removing internal links to these images, such as the slide label, they are no longer accessible via the viewing software, although the actual image data is still present. This approach is considered insufficient because identifying data can still be obtained with little effort and technical knowledge about the file structure. Level III, on the other hand, can be achieved by making the slide label physically unrecognizable or completely deleting the corresponding image from the WSI. This in turn makes it very difficult to trace back the slide to the patient since re-identification would require access to the issuing institution's information systems. Thus, level III anonymizations fulfill the requirements of the GDPR. Nevertheless, re-identification remains theoretically possible as the metadata in WSIs provides information about a) the slide or case itself and b) the capture modalities. In particular, the data collected during the digitization of WSIs, such as scanner serial numbers, acquisition dates and times, as well as operating system user names, can reveal a unique constellation that enables unambiguous traceability when the associated clinical systems can be accessed. Achieving Level IV, therefore, requires the removal of this sensitive metadata. In this way, the WSIs contain only relevant image data of the tissue and attributes such as dimensions, resolution, or structure.

However, it is already argued that the tissue itself can provide traceability to the corresponding case which is possible through modern image-matching algorithms with only a little computational effort. For example, within a multi-center study, it may be possible to identify the particular institute based on the specific staining. Assuming the disease is specific enough, the eligible patients can be reduced to a minimum. Thus, an individual patient can be identified by appropriate comparison between the WSI and the original slides. Holub et al. have published a preprint describing an attack scenario in which the attacker can access an identifying WSI dataset based on background information through feature extraction and similarity comparison of image details [18]. This thesis results in our final anonymization level V, in which we require the elimination of the spatial coherence of tissue sections to anonymize histopathological image data fully. To achieve this level of anonymity, data must not be retraceable to the original tissue based on the image content alone.

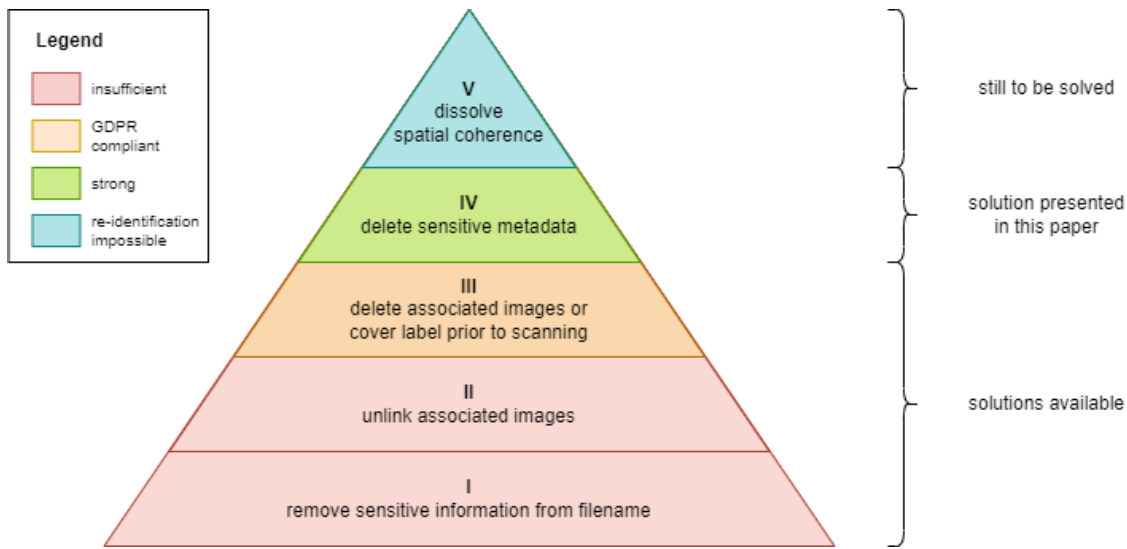

**Figure 2**: The five levels of WSI anonymization.

Various solutions for the three lower levels (I - III) have already been presented in the "Related work" section. Level I requires removing sensitive information from the file name, which can be solved via the operating system. The programmatic removal of the link to the label image and, if necessary, the macro image using the existing solutions [15, 16] is classified as Level II. If the label image is completely deleted or obscured before scanning, Level III is reached. To comply with the GDPR, it is generally sufficient to cover the labels before scanning. However, if sections are digitized in a diagnostic context, they would have to be scanned twice to obtain one file with the relevant information and a second anonymized file. Level IV requires that all sensitive metadata be deleted in addition to the label image. Levels III and IV can theoretically be achieved with conversion to DICOM or OMERO but require conversion to another file format, which in turn requires support for that additional file format. The software library presented in this paper closes this gap and enables level IV anonymization directly within the proprietary WSI formats. Therefore it is crucial that not only the associated images and identifying data are removed but also, depending on the manufacturer's format, all metadata that can provide additional information about the origin and the time when the WSI was acquired. Currently, there are no usable solutions for Level V anonymization. One possible approach is to dissipate the spatial coherence of the underlying tissue by irreversible tiling, making a direct comparison between WSI and the slide impossible. For this purpose, an abstraction layer between the virtual slide and the actual image data could be used as a proxy between the file format and an application or end user. Image tiles or regions are then retrieved based on tissue characteristics (such as staining or indication) and an arbitrary selection from a diverse collection of data. Such an approach, however, is particularly tailored to the needs of training and validation data for AI algorithms.

**Workflow**

The anonymization process consists of various individual steps, which are processed sequentially, as shown in Figure 3. First, the WSI is read and checked to see if the data format is supported. The file is then renamed, and a non-anonymized backup is created depending on the configuration. Then, the relevant tags and positions of the sensitive data are determined based on the structure of the respective data format. Finally, all relevant bytes are overwritten to ensure non-reproducibility.

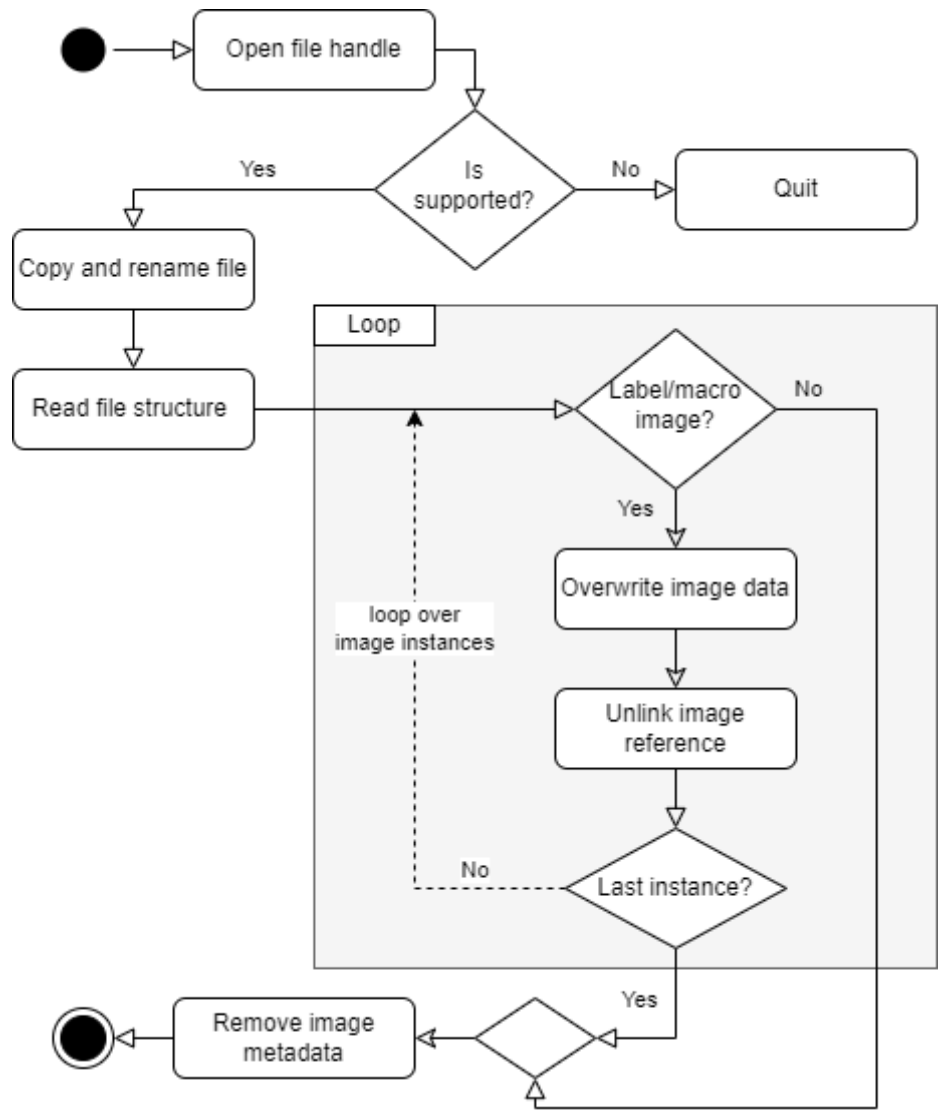

**Figure 3**: Generic workflow of the anonymization procedure.

For image data, it is best to set the corresponding binary data to zero, keeping in mind that the file format headers and the entire structure are preserved to maintain a valid file structure. Compressed image data, for instance, need to preserve the binary structure of the underlying compression algorithm stated in the WSI header (e.g. LZW, Deflate, or JPEG) to remain interpretable for external software. Similarly, that also applies to textual modifications within the file or directory headers. Textual metadata has, just as associated image data, fixed memory allocation that can be overwritten incorrectly, which would cause the whole file to become invalid. This requires carefully determining the offset pointers and lengths of the *data blobs* (binary large objects).

**Implementation**

The introduced anonymization library exposes two methods to a) determine the vendor/file format and b) execute the actual anonymization process. Within the anonymization method body, we implement the procedure described in the workflow section for each file format. Since quite a few formats implement their data structures based on the tiff specification (*Tagged Image File Format*), we encapsulate shared functionality for these formats. Tiff-based files are built on a standardized directory-based structure, namely *Image File Directories* (IFDs) [19]. The file as a whole, as well as each directory by itself, defines a header with file and directory-specific metadata, pointers referencing the binary data and finally, the subsequent tiff directory, creating a tightly connected chain of directories (see Figure 4). While most metadata can be accessed through common pre-defined tiff tags, some vendors decide to implement custom tags, exposing additional information related to the WSI. These vendor-specific tags and the non-standardized way of storing information within these tags complicate the anonymization process. Tiff files can be identified by the first few bytes, which indicate the endianness, the byte order of the data fields and the version of the file, followed by the first IFD offset. The initial IFD offset points to the first directory that starts with a two-byte entry count, defining a count of tiff tags following. The last tag entry within the directory header is again a pointer referencing the offset of the subsequent IFD. As tiff tags can only hold 12 bytes of data, these tags are usually used for tiny, partial information or as a reference to a data blob that does not fit inside the tag field. The dereferencing of the data is accomplished through a pointer, the length of the blob and further properties to describe the data structure. Image data includes, for example, the width and height of the image and further information about the color composition or compression. To allow for fast and easy navigation within the file format, the tiff file structure is parsed beforehand into an internal data structure with all relevant pointers and meta information.

For non-tiff-based file formats, the anonymization process is very similar. For instance, the Mirax file format provides a distinct metadata file, the *Slidedat.ini*, exposing information on the slide itself but also on the hierarchical structure of the WSI and the location of the associated binary image data. In contrast to tiff-based formats, the identification and removal are rather simple, as the information is represented as human-readable text. The removal of image data, however, is more complex, as the data is distributed over various binary files. Concrete offsets and blob lengths of the image data are stored in a further binary file, the Index.dat. These location parameters are parsed into an internal file representation to easily obtain pointer addresses and lengths for the relevant binary objects. Another aspect of the anonymization of Mirax files is the different format versions which in part differ in the location of the sensitive data, making it more complex to maintain a valid file structure.

Philips' iSyntax file format uses a different implementation approach. WSI-specific metadata, as well as the associated image data, are stored in the *Extensible Markup Language* (XML) header of the file, where the data is organized within hierarchical object nodes in the XML tree. The metadata contents are stored in the root node and accessed via the key of the specific attribute. These values are of predefined custom data types. While some data entries can simply be replaced by arbitrary values of the same data type, others are additionally restricted by constraints (e.g., an interval range). Furthermore, the image data of the macro and label image must be base64-encoded. The XML node that contains this data also consists of attributes that describe the representation of the image and may also need to be adjusted. [20]

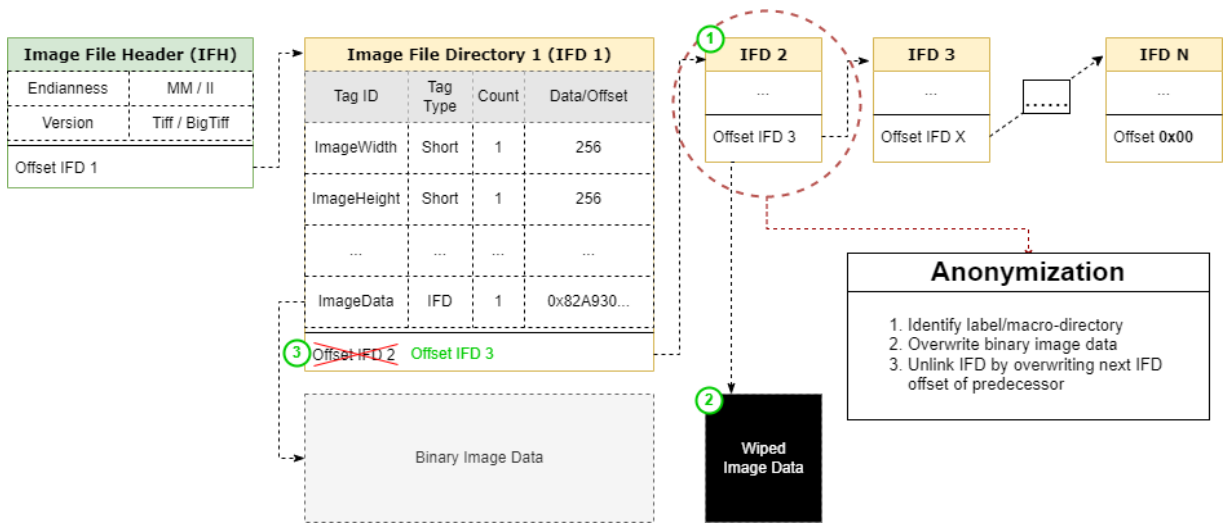

**Figure 4**: Structure of tiff-based file formats and the conceptual idea of overwriting and unlinking image data.

Deleting the associated image data is implemented in two separate steps. First, the image data is overwritten with a blank image so that it cannot be reconstructed later. Then, the image is unlinked so that the reference to the image data or its directory is removed. For instance, tiff-based files require an intact chain of subsequent directory references to remain a valid file structure that can be read by proprietary WSI-viewing software or SDKs and reverse-engineered libraries such as OpenSlide [21] and BioFormats [14]. To remain valid, the IFD pointer of the predecessor has to be overwritten with the IFD pointer of the ensuing directory or, in case it is the last directory, terminated with a null-pointer (see Figure 4). Incorrect assignment of memory addresses to the directory offsets will eventually result in dangling pointers and will thus corrupt the chain of referencing directories. For Mirax files, the unlinking is accomplished by rewriting the whole file structure within the *Slidedat.ini*, removing all references to the associated image data. Similarly, this applies to iSyntax files, where an intact XML structure is expected. It is not possible to simply remove individual object nodes, as this will invalidate the file structure that the proprietary Philips Pathology SDK expects. A further challenge is the potential compression method used for the image data. The compression method can be stated in the metadata or is assumed from the file format description, causing an invalidation of the structure when a divergent method is used.

Similar issues are encountered when anonymizing metadata. Usually, a fixed length of bytes is reserved at a certain pointer position or in a tiff tag entry. Exceeding the size limit of the preserved memory location can lead to format invalidation. As metadata is stored exclusively as a string data type, the sensitive data can not just be removed but has to be replaced by an arbitrary, content-free string of the same length as the original string.

## Results

With the presented library, the WSI file formats listed in Table 1 can be anonymized. Due to internal and structural software differences of scanners, it has to be ensured that the library performs equivalently and is well-tested on various slide formats.

Data from the OpenSlide WSI repository [22] and partners of the EMPAIA Consortium (EcosysteM for Pathology Diagnostics with AI Assistance) were used for implementation and testing. The image data and metadata described in the materials section and referenced in Appendix A are deleted and unlinked without invalidating the native file format. Thus, the resulting WSI files are still usable by the designated proprietary or common open-source software. The total runtime for most formats is less than a second on an ordinary workstation and, therefore, a neglectable benchmarking factor. The software is initially implemented to support all common file formats of scanners used in clinical routine, including Leica, Hamamatsu, 3DHistech, Roche and Philips. It is open for extensions and provides a modular implementation interface to ease the integration of additional proprietary file formats. Capabilities for reading and writing tiff file structures or manipulating image data and strings can further enhance the development process.

| Vendor | File format extension | Scanner types | Comment |
| --- | --- | --- | --- |
| 3DHistech | *.mrxs | - P150<br>- P250<br>- P1000 | |
| Hamamatsu | *.ndpi | - XR<br>- XT2<br>- S360 | Label and macro image are one image and cannot be deleted separately. |
| Leica / Aperio | *.svs<br>*.tif | - AT20<br>- GT450 | |
| Philips | *.isyntax | - IntelliSite Ultra Fast Scanner | |
| Roche / Ventana | *.bif<br>*.tif | - VS200<br>- iScan Coreo | Label and macro image are one image and cannot be deleted separately. |

**Table 1**: Supported slide scanner and file formats.

The presented library is implemented in C, as this allows file access on the operating system level and thus requires almost no external dependencies for the anonymization process. Hence, it facilitates the development, usage and accessibility of the anonymization library. Seek, read and write operations are performed time-efficiently and are portable from and to various operating systems without further adjustments. Due to the lightweight architecture of the core library, the C code is easily callable from C++ code and can be wrapped for different programming languages. As a standalone tool, the software can be compiled into a Command Line Interface (CLI) application that allows for basic parametrization of the anonymization process. For instance, users can specify via CLI-flag, if the macro image should be preserved or if the associated images should solely be overwritten or additionally unlinked. Since the anonymization of WSIs plays an important role for uploading in our cloud-hosted EMPAIA ecosystem, the library is demanded to expose its interfaces to the frontend JavaScript Code, enabling client-side execution of the code within the browser.

Through WebAssembly (WASM), an assembly-like language that provides a compilation target for languages like C, C++ or Rust, the C code can be executed within the browser. Read- and write operations are ported from a file-based to a stream-based way, where successive data chunks are expected as input rather than the entire file. The implementation and compilation are done with emscripten, an LLVM/Clang-based compiler optimized for C and C++ source code. According to the developers, the execution time is about half the time of the native speed, resulting in a negligible increase in computation time [23]. The WASM-compiled library is available within the project's package registry, allowing the download and installation of the JS package via the package-manager *npm*. The source code is publically available on Gitlab.com[1] as part of the EMPAIA group.

## Discussion

Depending on the country of origin, we are confronted with different legal requirements and regulations for WSI anonymization, which particularly affect the partners involved in the data exchange and the utilized communication channels. Furthermore, it is also advisable, from an ethical point of view, to delete all patient-related information from the data. Since such functionality does not exist in the scanning software of the various vendors and most of them only supply proprietary WSI formats, a broadly applicable solution for anonymization is essential. Attempts to standardize WSIs such as DICOM WSI or OME.TIFF are major steps toward independence from proprietary formats and will considerably simplify the identification and deletion of sensitive data. In particular, the standardization of DICOM eliminates dependencies on scanner manufacturers by providing a conceptual framework for a uniform definition of data fields. Despite the very active conception and development of the DICOM standard for digital pathology, the acceptance and use of the open format specification are still rudimentary today. However, this is not so much due to the technical maturity of the standard but rather to the fact that manufacturers of scanners and image management/archiving systems are slow to implement it and integrate it into their existing solutions [24, 25]. In addition, the adoption of DICOM WSI as a primary output of slide scanners is not yet comprehensively accepted or even implemented. Converting WSIs to other formats for anonymization purposes is neither practicable nor is it always possible.

We have defined an anonymization policy with five distinct levels of anonymization and evaluated their implications on data privacy and protection and GDPR compliance. With the presented software library, a level IV anonymization can be achieved that is GDPR compliant and additionally makes re-identification based on the metadata impossible. However, for level V anonymization, the appropriate concepts must first be developed and implemented. It must be impossible to assign image data to the original in a direct comparison. For this purpose, it is at least necessary to dissolve the spatial context of the image data. One way to achieve this is to subdivide the WSI into individual image tiles that can no longer be combined into coherent tissue. This is achieved when no mutual information is contained in the tiles by excluding overlapping areas and prominent edges. Web services are a common approach to serve as an abstraction layer between file formats and image data, providing only image-specific metadata and plain pixel information without any knowledge of the underlying WSI. Picture Archiving and Communications Systems (PACS), Labor Information Systems (LIS), or other Image Management Systems (IMS) can then be charged for image sourcing. At the same time, the web service defines a communication interface to handle requests from applications or end-users to these underlying systems. For instance, queries could be made for a specific combination of tissue, stain, analysis or indication, and a set of non-cohesive image tiles would be returned as a result. It should be noted, however, that such a concept undermines the classical use of WSIs in digital pathology since the usage of these image data is very limited to specific applications that do not bother the missing spatial information, such as training and validation data for artificial intelligence. The opposite approach of swarm or federated learning, on the other hand, does not seek to share image data over public networks but rather to distribute the algorithms to data nodes in a decentralized way using distributed/edge computing and blockchain technology. With this approach, data privacy and security concerns need not be addressed further [26]. Again, it should be noted that this approach is primarily suited for ML/AI training and additionally requires a complex deployment strategy at the respective data nodes (e.g., institutions), since in this scenario not only the data but also the algorithms must be secured against unauthorized access [27].

---
[1] https://gitlab.com/empaia/integration/wsi-anon

The library we present allows for the anonymization of WSIs to be performed while retaining the original, proprietary data format. Furthermore, the communication channel is also addressed by providing a WSI upload module that integrates anonymization into the upload process, ensuring that sensitive data never leaves the uploading system. Despite providing a flexible solution for anonymization, it is still required to integrate an external solution into an existing pathology workflow. The preferred way would be built-in functionality in the proprietary scanning or archiving software that has not yet been provided by vendors so far. This would also eradicate potential uncertainties regarding the format descriptions and enable the joint use of slides for a purpose other than routine diagnostics. This dual use of WSIs in routine diagnostic work, research or education needs to be tackled, as patient-related data is still required for clinical use cases. In addition, the tool requires continuous updates since the proprietary WSI formats are also constantly being revised or new formats/format versions are being developed, which then need to be incorporated into the tool. For Leicas' GT450 scanner, for example, the previous WSI format was migrated from ordinary Tiff to BigTiff to enable WSIs above 4 GB size, resulting in a breaking change of the file structure.

If subsequent pseudonymization is desired, a more comprehensive workflow must be implemented, including the integration of identity and access management systems so that data can be traced back to the respective patients. There are already available solutions for this purpose of pseudonymization of healthcare data. The MOSAIC Project, for example, offers a modular approach of several independent tools dedicated to pseudonymization and the establishment of trust broker and ID management systems for medical data records [28]. However, even for using ID management and pseudonymization software, it is a prerequisite that WSIs can be fully anonymized and associated images can be replaced with an artificial image containing pseudonymized information.

## Conclusions

Both legal requirements and ethical considerations in general demand a secure solution to erase sensitive data from WSIs. We presented an anonymization policy compliant with the GDPR and defined different levels of privacy and data security for Whole Slide Imaging. An examination of proprietary WSI formats has shown that file names may contain critical data, in additional image data and the underlying metadata. Accordingly, the ability to delete this data is an indispensable requirement for working with WSIs. Beyond that, it is useful to be able to replace the relevant data, for example, to enable pseudonymization.

Our tool can be used to remove associated images as well as relevant metadata in the formats of 3DHistech, Hamamatsu, Leica, Philips and Roche and is thus compliant with level IV of our anonymization policy. The tool itself can be downloaded and executed directly as a CLI tool. Additionally, the C library source code is distributed under the MIT license with bindings for python and JavaScript to allow easy integration into other software components. Thus, exchanging WSIs is possible regardless of the legal situation and the respective partners involved.

# Declarations

Conflicting interests: The authors declare that they have no known competing financial interests or personal relationships that could have appeared to influence the work reported in this paper.

Funding: This work was funded by the German Federal Ministry for Economic Affairs and Climate Action (BMWK) [Grants 01MK20002A, 01MK20002B].

Ethical approval: This article does not contain any studies with human participants or animals performed by any of the authors.

Guarantor: TB

Contributorship: TB and MF researched literature and wrote the manuscript; MF led the implementation and technical analysis; TB developed the anonymization policy and led the legal analyses; ID, DR and CJ were involved in the software implementation; PH and NZ initiated the work on this topic, gave extensive feedback to both the technical as well as the legal aspects and were involved in communicating with the scanner vendors. All authors reviewed and edited the manuscript and approved the final version of the manuscript.

Acknowledgement: We would also like to thank Dr. Rasmus Kiehl and Dr. André Homeyer for their comments that greatly improved the manuscript.

# Appendix A

| | Leica/Aperio | Hamamatsu | 3DHistech/Mirax | … |
|---|---|---|---|---|
| Structure | Tiff/BigTiff | Tiff | Configuration file | |
| Label | x | - | x | |
| Macro | x | x | x | |
| Metadata | ScanScope ID<br>Date<br>Time<br>User<br>Filename | Macro.S/N<br>NDP.S/N<br>Created<br>Updated | SLIDE_NAME<br>PROJECT_NAME<br>SLIDE_ID<br>SLIDE_CREATIONDATETIME<br>SCANNER_HARDWARE_ID<br>SLIDE_UTC_CREATIONDATETIME<br>ProfileName | |

| | … | Roche/Ventana | Philips |
|---|---|---|---|
| Structure | | Tiff/XML | XML |
| Label | | x | x |
| Macro | | - | x |
| Metadata | | JP2FileName<br>UnitNumber<br>UserName<br>Barcode1D<br>Barcode2D<br>BaseName<br>BuildDate | DICOM_ACQUISITION_DATETIME<br>DICOM_DEVICE_SERIAL_NUMBER<br>PIIM_DP_SCANNER_OPERATOR_ID<br>PIM_DP_UFS_BARCODE<br>PIIM_DP_SCANNER_RACK_NUMBER<br>PIIM_DP_SCANNER_SLOT_NUMBER |